\documentclass[twocolumn,prb,showpacs,preprintnumbers,amsmath,amssymb]{revtex4}

\usepackage{graphicx} 
\usepackage{dsfont} 

\begin{document}


\title{Polarization sensitive spectroscopy of charged Quantum Dots}
\author{E. Poem}
\email{poem@tx.technion.ac.il}
\author{J. Shemesh}
\author{I. Marderfeld}
\author{D. Galushko}
\author{N. Akopian}
\author{D. Gershoni}
\affiliation{Department of physics, The Technion - Israel institute of technology, Haifa, 32000, Israel}
\author{B. D. Gerardot}
\author{A. Badolato}
\author{P. M. Petroff}
\affiliation{Materials Department, University of California Santa Barbara, CA, 93106, USA}

\date{\today}

\begin{abstract}
We present an experimental and theoretical study of the polarized
photoluminescence spectrum of single semiconductor quantum dots in
various charge states. We compare our high resolution polarization
sensitive spectral measurements with a new many-carrier theoretical
model, which was developed for this purpose. The model considers
both the isotropic and anisotropic exchange interactions between all
participating electron-hole pairs. With this addition, we calculate
both the energies and polarizations of all optical transitions
between collective, quantum dot confined charge carrier states. We
succeed in identifying most of the measured spectral lines. In
particular, the lines resulting from singly-, doubly- and triply-
negatively charged excitons and biexcitons. We demonstrate that
lines emanating from evenly charged states are linearly polarized.
Their polarization direction does not necessarily coincide with the
traditional crystallographic direction. It depends on the shells of
the single carriers, which participate in the recombination process.
\end{abstract}


\maketitle

\section{Introduction}
Quantum dots (QDs) are nano-structures, which confine electrons and
holes in all 3 dimensions. This confinement results in a discrete
spectrum of single carrier energy levels and spectrally sharp
optical transitions between them. The photoluminescence (PL)
spectrum of single self-assembled semiconductor QDs is usually
composed of many discrete spectral lines. The variety of lines
originates from optical transitions between various many carrier
configurations and different QD charge
states \cite{hartmann,dekel00,finley,urbazek,seguin}.\\
Several experimental techniques are used for identifying a given
spectral line by associating it with a specific optical transition.
These techniques include excitation intensity dependent PL
spectroscopy, which distinguishes between single-exciton and
multi-exciton transitions \cite{regelman1} and second order
intensity cross-correlation measurements, which determines the
temporal sequence by which the emission occurs in
general\cite{regelman1}, and identifies radiative cascades in
particular \cite{nika_ent, monroe}. PL excitation (PLE)
\cite{finley,ware} as well as electro\cite{warburtonNAT}- and
magneto\cite{bayer_rev,shlomi}-PL spectroscopies are used to further
provide information regarding the QD's charge state during the
optical transitions.\\
Unfortunately, even when all of these methods are combined,
occasionally, some lines still remain unidentified\cite{besombeprl}.\\
Polarization sensitive PL and PLE spectroscopy have also been
applied to aid in line identification. Most notably, the neutral
exciton and neutral biexciton lines are split into two cross
linearly polarized doublets\cite{gammon,kulakovskii,woggon}, while
singly charged excitonic lines are unpolarized, and display large
circular polarization memory \cite{ware,warburtonAPL,marie} when
excited
quasi-resonantly.\\
In this work we focus our studies on polarization sensitive PL
spectroscopy of single semiconductor quantum dots. We carefully
measure the polarization of the PL spectra under various excitation
conditions. Our results are then compared with, and analyzed by, a
novel theoretical many charge-carriers model. The method used for
the calculation of the many-carrier states and optical transitions
between them is the full-configuration-interaction (FCI) method
\cite{barenco}. The novelty in our model is in its inclusion of the
electron-hole exchange interaction
(EHEI)\cite{ivch_pikus,bayer_rev,akimov}. We show that the model
provides a very good understanding of the experimental measurements.\\
The manuscript is organized as follows: In section \ref{sec:exp} we
describe the sample and the experimental setup used for the
polarization sensitive PL spectroscopy. In section \ref{sec:theo} we
describe the theoretical model and in section \ref{sec:comp} we
compare theoretical and experimental results. A short summary is
given in section~\ref{sec:sum}.
\section{\label{sec:exp}Experimental methods}
\subsection{Sample}The sample was grown by molecular beam epitaxy on a [001]
oriented GaAs substrate. One layer of strain-induced InGaAs QDs was
deposited in the center of a one wavelength GaAs spacer layer. The
height and composition of the QDs were controlled by partially
covering the InAs QDs by a 30\AA\ thick layer of GaAs and by
subsequent 30 seconds growth interruption\cite{garcia} to allow
diffusion of In(Ga) atoms from (into) the strained islands. The
growth resulted in ${\rm In}_{{\rm x}}{\rm Ga}_{1-{\rm x}}{\rm As}$
QDs whose exact shape, lateral size, composition and strain profile
are unknown.\\
The sample was not rotated during the growth of the strained layer,
resulting in a variable density of QDs across the sample's
surface\cite{regelman1}. The estimated density of QDs in the areas
that were measured is $10^8 {\rm cm}^{-2}$.\\
The optical microcavity was formed by distributed Bragg reflecting
(DBR) stacks of 25 and 11 periods of alternating AlAs/GaAs quarter
wavelength layers below and above a GaAs spacer layer, respectively,
giving a Q-factor of $\sim$500. 
The spacer layer was grown to a width close to the wavelength in
matter of the light emitted from the QDs due to ground state
\mbox{e-h} pair recombinations (1~$\lambda$~cavity). The microcavity
improves the efficiency of photon collection, but limits the energy
in which photon collection is possible. In particular, emission of
photons with energies smaller than the microcavity mode energy is
forbidden. Therefore, the density of QDs which emit efficiently is
roughly two orders of magnitude lower than their actual density
\cite{ramon}. In order to electrically charge the QDs, a p-i-n
junction was formed by n-doping the substrate and the bottom DBR and
p-doping the top DBR, while leaving the GaAs spacer intrinsic. An
extra AlAs barrier was grown inside the GaAs spacer between the
p-type region and the QDs. This barrier prolongs the hole's
tunneling time into the QDs at forward bias and out of them at
reverse bias, with respect to the tunneling time of the electron.
This enables negative charging upon forward bias and positive
charging upon reverse bias.\\
The top electrical contact of the sample was made of a
semi-transparent layer of indium-tin oxide in order to provide
optical accessibility. The sample was not patterned or processed
laterally to prevent obscuration of the QD emission and its
polarization.
\subsection{Optical characterization}
\begin{figure}[tbh]
\includegraphics[width=0.5\textwidth]{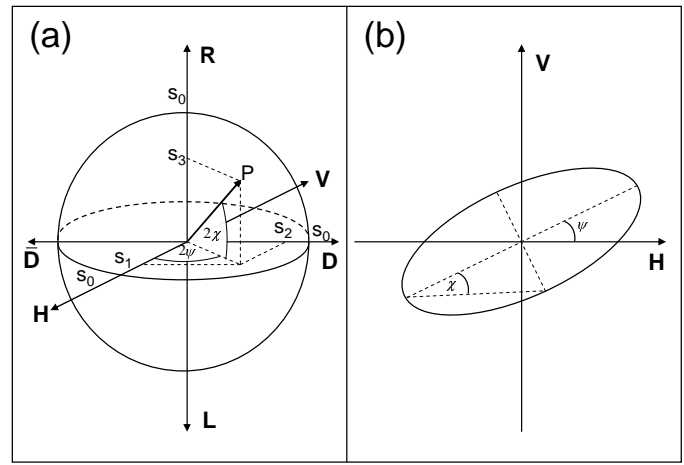}
\caption{\label{fig:poincare} (a) The polarization state represented
as a vector P on the Poincar\'e sphere. (b) The same polarization
state represented as the loci of points that the electric field of
the light obtains, during one period, in a plane perpendicular to
its propagation direction. $s_{0..3}$, are the experimentally
determined four Stokes coefficients\cite{jackson}.}
\end{figure}

\begin{figure}[tbh]
\includegraphics[width=0.5\textwidth]{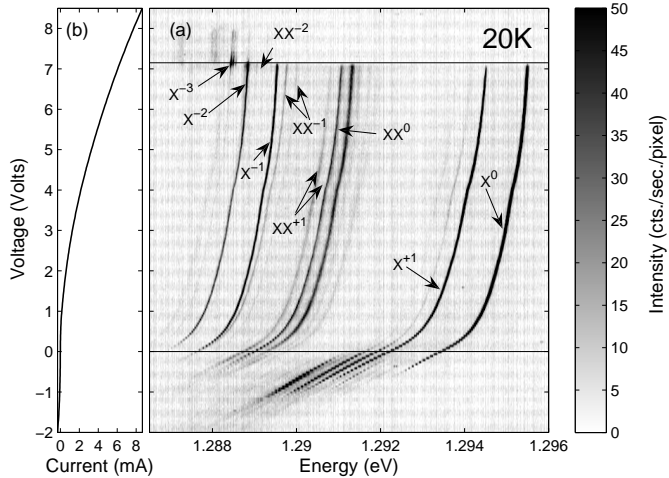}
\caption{\label{fig:PLV} (a) Measured PL spectra from a single SAQD,
as a function of the bias on the device. The QD was excited by 1.47
eV cw laser light. The various spectral lines are labeled by X (XX)
for single (double) initial e-h pair occupation and a superscript
which denotes the QD charge state during the recombination. The
horizontal solid lines mark the voltages for which spectra are
presented in Fig.~\ref{fig:PL_polar}. (b) The current through the
device as a function of the bias voltage.}
\end{figure}

For the optical measurements we used a diffraction limited low
temperature confocal optical microscope \cite{dekel98, dekel00}. The
sample was mounted on a cold finger of a He-flow cryostat,
maintaining temperature of about $\sim$20K.  A X60 in-situ
microscope objective was used in order to focus cw or pulsed laser
light at normal incidence on the sample surface. The emitted light
was collected by the same microscope objective. The objective was
accurately manipulated in three directions using computer-controlled
motors. The collected light was spatially filtered, dispersed by a
1~meter monochromator and detected by a nitrogen-cooled CCD array
detector. The system provides diffraction-limited spatial
resolution, both in the excitation and the detection channels and
spectral resolution of about 15 $\mu eV$ per one CCD camera pixel.\\
The polarization of the emitted light was analyzed by two computer
controlled liquid crystal variable retarders and a linear polarizer
in-front of the monochromator. The degree of polarization of the
emitted light and its polarization state were deduced by six
independent measurements of differently polarized spectra and
calculation of the Stokes parameters \cite{jackson}. Throughout this
work we use the symbol H (V) for linear light polarization along the
[1\=10] ([110]) crystallographic axis of the sample. These in-plane
orientations are determined by cleaving. The symbol
D=$\frac{1}{\sqrt{2}}$(H$+$V) (\=D=$\frac{1}{\sqrt{2}}$(H--V)) is
used for the 45$^{\circ}$ (-45$^{\circ}$) diagonal polarization,
while the symbol R=$\frac{1}{\sqrt{2}}$(H$+i$V)
(L=$\frac{1}{\sqrt{2}}$(H--$i$V)) is used for the right
(left) hand circular polarization.\\
A general state of polarization can be represented as a vector
inside the Poincar\'e sphere. Figure \ref{fig:poincare} shows a
vector in the Poincar\'e sphere and its relation to the shape and
orientation of the polarization.
\\
In Fig.~\ref{fig:PLV} we show the PL spectra from a single QD as a
function of the voltage applied to the sample. The QD was excited by
a cw 1.47~eV Ti-sapphire laser light. The current through the device
as a function of the voltage is also shown.
\\
The specific structure of our sample is such that at forward biases
(above $\sim$7~volts) the QDs are negatively charged as clearly evident by the abrupt
step in the emission energy. This injection induced charging mechanism is similar
to that reported earlier\cite{urbazek,warburtonNAT}. At large reverse biases, however,
the QD is increasingly positively charged, due to vast differences between
the tunneling-out rates of electrons and holes\cite{wareproceedings, ware}.\\
The spectral line identification in Fig.~\ref{fig:PLV} is based on
the order by which the lines appear and disappear as the voltage on
the device increases. Information gained from excitation intensity
dependence PL spectroscopy (not shown) and polarization sensitive
spectroscopy (see below) is also used for this purpose.\\
In Fig.~\ref{fig:PL_polar} we present the measured polarization
sensitive spectra for the bias voltages indicated by horizontal
lines in Fig.~\ref{fig:PLV}.
\begin{figure}[tbh]
\includegraphics[width=0.5\textwidth]{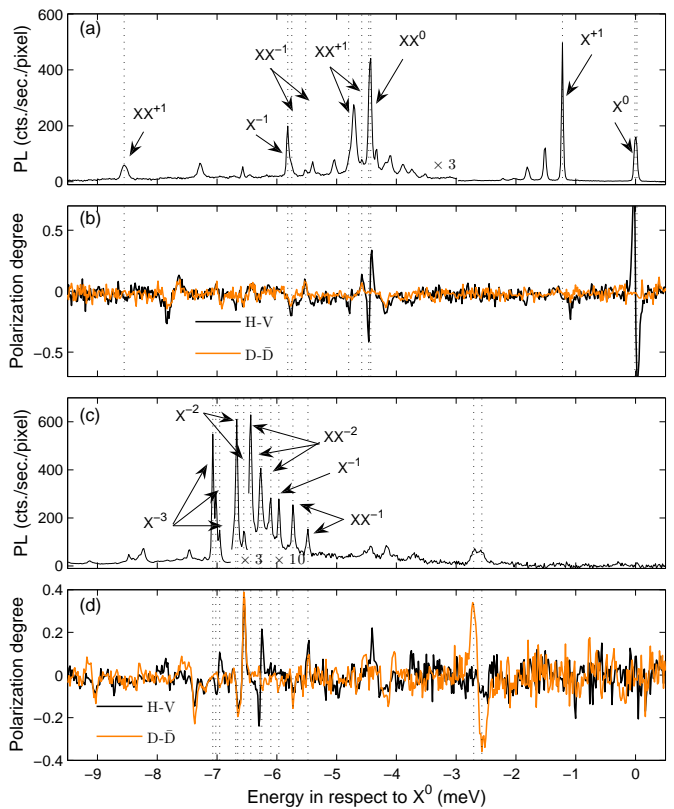}
\caption{\label{fig:PL_polar}(color online) (a) and (c) Measured PL
spectrum for bias voltage of 0V and 7.15V, respectively. The energy
is measured from the energy of the $X^0$ line. (b) and (d) PL
polarization spectra for bias voltage of 0V and 7.15V, respectively.
The black (orange) lines present the polarization as projected on
the H--V (D-\=D) axis of the Poincar\'e sphere. Vertical dash lines
at various spectral lines are drawn to guide the eye.}
\end{figure}\\
We note here that the spectral shapes of the observed negatively
charged lines and in particular the fine structure components of
X$^{-2}$, XX$^{-2}$
and X$^{-3}$ are similar to those observed also in previous
works\cite{urbazek,warburton_zunger}.\\
In Fig.~\ref{fig:PL_polar}(a), the QD was on average neutral. The
neutral, as well as the singly negatively and singly positively
charged exciton and biexciton spectral lines are observed. The
corresponding polarization spectra projected on the linear H--V and
on the linear D-\=D axes of the poincar\'e sphere are shown in
Fig.~\ref{fig:PL_polar}(b). The projections are calculated by
subtracting the two cross-linearly polarized spectra dividing by
their sums. Division by zero is avoided by adding a constant equals
to the standard deviation of the background noise to each spectrum.
The projection on the R--L axis of the Poincar\'e sphere was zero to
within our experimental uncertainty (not shown). From these two
projections, the actual magnitude and direction of the linearly
polarized lines can be straightforwardly determined. In
Figs.~\ref{fig:PL_polar}(c) and \ref{fig:PL_polar}(d) we present the
PL spectrum and its linear polarization projections, respectively,
for a bias voltage of 7.15 volts at which the QD was on average,
negatively charged with two to three electrons.
\section{\label{sec:theo}Theoretical Model}
The model that we developed is a relatively simple many-carrier
model which includes the electrostatic interactions between the QD
confined charge carriers. Unlike previous, similar
models\cite{barenco,dekel00,seguin}, which neglected the electron-hole
exchange interaction (EHEI), our model includes it. This interaction
is indeed orders of magnitude smaller than the direct Columbic
terms. Spectrally, it is only significant when the fine excitonic
structure of the spectrum is considered. However, when the
polarization spectrum is considered, this anisotropic
interaction\cite{ivch_pikus,ivchenko,takagahara} is by far the
leading term.\\
Our model is constructed as follows:\\
We first solve a single carrier problem for the electron and for the
hole in the QD. In this manner, we obtain a consistent set of single
charge carrier energies and associated wavefunctions.\\
We then use this set of energies and wavefunctions in order to
construct a many-carrier second quantization Hamiltonian, which
includes the electrostatic interaction between the confined
carriers. In particular we consider the EHEI which is introduced
into our model using a semi-phenomenological approach.\\
The many body Hamiltonian is then diagonalized, thus obtaining the
collective many carriers energies and wavefunctions. We then use the
dipole approximation to calculate the optical transitions between
the many carriers states for a given light polarization. From these
calculations we finally construct the polarization sensitive
emission spectrum, which is used for comparison with the
experimental measurements.
\subsection{The single-carrier problem}
The single-carrier energies and wavefunctions are calculated using
the slowly varying envelope function approximation
(SVEFA)~\cite{cardona}. We use one (doubly Kramers degenerate) band
for the electron and one band for the hole without band mixing (`One-band SVEFA').
This approximation results in two independent Schr\"{o}dinger
equations for the envelope functions of the electron and that of the
hole. The potential of the QD is approximated by a finite three
dimensional potential well in the form of a rectangular slab, with
the long (short) side oriented along the H (V) direction, and with
different dimensions and offsets for the two types of carriers. The
parameters that we used are listed in table \ref{tab:QD_params}. We
solved the differential equations numerically, using the finite
differences method, thus obtaining the single particle
eigen-energies and envelope wavefunctions.
\begin{table}[tbh]
\caption{\label{tab:QD_params} The QD parameters used to calculate the
single-carrier energies and envelope wavefunctions.}
\begin{ruledtabular}
\begin{tabular}{l|c|c}
Parameter&Value&Units\\
\hline
\hline
QD shape & Rectangular slab & - \\
QD size for the electron &  & \\
(Length x Width x Height) & 244 x 232 x 34 & \AA \\
Hole/Electron length ratio & 0.72 & - \\
Electron effective mass & 0.065\cite{dudi_8kp} & $m_0$ \\
Hole z-direction effective mass & 0.34\cite{dudi_8kp}& $m_0$ \\
Hole in-plane effective mass & 0.25 & $m_0$ \\
Electron potential offset & -324 & $m$eV \\
Hole potential offset & -108 & $m$eV \\
GaAs band gap & 1.519\cite{landt_born} & eV \\
Band gap of QD material & 1.087 & eV\\
In$_{0.5}$Ga$_{0.5}$As dielectric constant & 13.8\cite{landt_born} & - \\
E$_p$ of In$_{0.5}$Ga$_{0.5}$As & 25.5\cite{dudi_8kp} & eV
\end{tabular}
\end{ruledtabular}
\end{table}
\subsection{The many-carrier Hamiltonian}
The second quantization many-carrier Hamiltonian for the QD
containing both electrons and holes is given
by\cite{dekel00,barenco}:
\begin{equation}\label{eq:H_general_form}
\hat{H}=\hat{H}_0+\hat{H}_{ee}+\hat{H}_{hh}+\hat{H}_{eh}
\end{equation}where $\hat{H}_0$ is the single carrier Hamiltonian, $\hat{H}_{ee}$ ($\hat{H}_{hh}$) is the electron-electron (hole-hole) interaction Hamiltonians, and 
\begin{equation}\label{eq:Heh}
\hat{H}_{eh}=\sum_{i_1,i_4,j_2,j_3}{(-C^{ehhe}_{i_1,j_2,j_3,i_4}+C^{hehe}_{j_2,i_1,j_3,i_4})\hat{a}^{\dag}_{i_1}\hat{b}^{\dag}_{j_2}\hat{b}_{j_3}\hat{a}_{i_4}}
\end{equation}
is the electron-hole interaction Hamiltonian.
The electron creation operator $\hat{a}^{\dag}_{i}$ and the hole
creation operator
$\hat{b}^{\dag}_{j}$ seperately satisfy the regular Fermionic anti-commutation relations.\\
The quantities of the form $C^{p_1p_2p_3p_4}_{n_1,n_2,n_3,n_4}$,
where $p_{1..4}$ can be either `\textit{e}' (for electron) or
`\textit{h}' (for hole), and the indices $n_{1..4}$ run
over the appropriate states, are the Coulomb interaction integrals.\\
Computation is only feasible with limited number of single carrier
states. Therefore, only the first 12 lowest energy electron and hole
states are usually considered in our calculations.\\
For the single-carrier wavefunctions which are calculated using the
SVEFA, the Coulomb integrals can be separated into long-range
(inter-unit-cell), and short-range (intra-unit-cell) integrals.
The long-range integral can be expanded into a Taylor series in
$\vec{r}_1-\vec{r}_2-(\vec{R}_1-\vec{R}_2)$, where $\vec{R}_{1(2)}$
is the lattice vector nearest to the position vector
$\vec{r}_{1(2)}$. In most cases the zeroth order of that series is
much larger than other orders and also larger than the short range
integral \cite{cardona}. For the one band SVEFA,
the zeroth order of the
long-range interaction is given by \cite{cardona}:\\
\begin{widetext}
\begin{equation}\label{eq:Coulomb_EFA}
C^{p_1p_2p_3p_4}_{n_1,n_2,n_3,n_4}=\delta_{p_1,p_4}\delta_{p_2,p_3}\delta_{S_{n_1},S_{n_4}}\delta_{S_{n_2},S_{n_3}}\int{\int{d^3r_1d^3r_2\phi^{p_1*}_{n_1}(\vec{r}_1)\phi^{p_2*}_{n_2}(\vec{r}_2)\frac{\textsl{e}^2}{\epsilon|\vec{r}_1-\vec{r}_2|}\phi^{p_2}_{n_3}(\vec{r}_2)\phi^{p_1}_{n_4}(\vec{r}_1)}}\end{equation}
\end{widetext}
where $\phi^{p}_{n}(\vec{r})$ is the envelope function of the $n^{th}$ state of a single carrier of type $p$, $\textsl{e}$ is the electron charge, $\epsilon$ is the dielectric constant of the QD material, and $S_n$ is the (pseudo) spin of state $n$.\\
The elliptic disk shape of our model QD is symmetric under
reflections about planes perpendicular to its main symmetry axes.
Therefore, single carrier envelope functions are either odd or even
under these reflections. The term $\frac{1}{|\vec{r}_1-\vec{r}_2|}$
is even under the application of the same reflection for both
$\vec{r}_1$ and $\vec{r}_2$. Therefore, the parity of the integrand
in Eq.~(\ref{eq:Coulomb_EFA}) under such a `double reflection' is
determined by the parities of the envelope functions only. Whenever
the integrand is odd under a `double reflection', the integral
vanishes. We use these symmetry considerations in order to reduce
the required computation resources.
\subsection{Electron-hole exchange interaction}
The zeroth order term in the long range EHEI ,
$C^{hehe}_{j_2,i_1,j_3,i_4}$ (Eq.~(\ref{eq:Coulomb_EFA})) equals
zero. Therefore, higher order terms in the long-range  as well as
the short-range exchange integral must be
considered\cite{takagahara, gupalov}.\\
The pseudo-spin structure of the EHEI for the lowest energy envelope
functions is deduced from symmetry considerations (the method of
invariants)\cite{ivch_pikus, bayer_rev}. The SVEFA requires that the
same considerations hold also for any other combination of envelope
functions\cite{takagahara}. Thus, we express the
electron-hole-exchange terms $C^{hehe}_{j_2,i_1,j_3,i_4}$ as
follows:
\begin{widetext}
\begin{equation}\label{eq:Ehx_Inv}
C^{hehe}_{j_2,i_1,j_3,i_4}=\frac{1}{2}\left(
\begin{array}{cccc}
\Delta^{nj_2,ni_1,nj_3,ni_4}_0 & \Delta^{nj_2,ni_1,nj_3,ni_4}_1 & 0 & 0 \\
\Delta^{nj_2,ni_1,nj_3,ni_4*}_1 & \Delta^{nj_2,ni_1,nj_3,ni_4}_0 & 0 & 0 \\
0 & 0 & -\Delta^{nj_2,ni_1,nj_3,ni_4}_0 & \Delta^{nj_2,ni_1,nj_3,ni_4}_2 \\
0 & 0 & \Delta^{nj_2,ni_1,nj_3,ni_4*}_2 & -\Delta^{nj_2,ni_1,nj_3,ni_4}_0
\end{array}\right)
\end{equation}
\end{widetext}
Where $ni_k$ represents the index of the envelope function belonging
to state number $i_k$. The e-h pseudo spin base for the matrix are
the functions: \mbox{\{$|\downarrow\Uparrow\rangle$,
$|\uparrow\Downarrow\rangle$,
$|\uparrow\Uparrow\rangle$, $|\downarrow\Downarrow\rangle$ \}.}\\
The terms $\Delta_0$ and $\Delta_2$ are mainly affected by the
short-range interaction\cite{ivch_pikus,takagahara}. This
intra-unit-cell interaction is not sensitive to the details of the
slowly varying envelope wavefunctions\cite{takagahara}. Therefore,
we assume that all the non-vanishing $\Delta_0^{n_2,n_1,n_3,n_4}$
and $\Delta_2^{n_2,n_1,n_3,n_4}$ terms have the same values,
$\Delta_0$ and $\Delta_2$, respectively. The values that we used
were chosen such that the calculated X$^0$ spectrum would fit the
magneto-PL measured X$^0$ spectrum\cite{shlomi}. Since the
short-range interaction is even under `double
reflections'\cite{takagahara}, the symmetry considerations which aid
in identifying the vanishing Coulomb integrals apply also in
identifying the
vanishing EHEI terms $\Delta_0$ and $\Delta_2$.\\
The $\Delta_1^{n_2,n_1,n_3,n_4}$ integrals are mainly affected by
the second order terms in the expansion of the long-range
interaction, which are given by \cite{ivch_pikus, takagahara}
\begin{widetext}
\begin{equation}
\label{eq:Ehx_Tak}\frac{1}{2}\Delta_1^{n_2,n_1,n_3,n_4}=\int\int\phi_h^{n_2*}(\vec{r}_1)\phi_e^{n_1*}(\vec{r}_2)\frac{\textsl{e}^2\vec{\mu}^{\dagger}_{\downarrow,\Uparrow}(\mathds{1}-3\hat{n}\hat{n}^{\dagger})\vec{\mu}_{\uparrow,\Downarrow}}{\epsilon|\vec{r}_1-\vec{r}_2|^3}\phi_h^{n_3}(\vec{r}_2)\phi_e^{n_4}(\vec{r}_1)d^3r_1d^3r_2
\end{equation}
\end{widetext}
where $\hat{n}$ is a unit vector in the direction of
$\vec{r}_1-\vec{r}_2$, $\mathds{1}$~is the 3$\times$3 unit matrix,
and $\vec{\mu}_{\uparrow,\Downarrow}$ is the valence-conduction band
dipole matrix element.  The dipole matrix element ($\vec{\mu}$) is
related to the momentum matrix element ($\vec{M}$) through the
particle's mass and the energy difference between the dipole
states\cite{in_cohen_tanudji} (the bandgap energy $E_g$),
\begin{equation}\label{eq:dipole_momentum}
\vec{\mu}_{\uparrow,\Downarrow
(\downarrow,\Uparrow)}=\frac{-i\hbar}{m_0E_g}\vec{M}_{\uparrow,\Downarrow(\downarrow,\Uparrow)}
\end{equation}
where the conduction-valence band momentum matrix elements are given
by,\cite{ivchenko,dudi_8kp}
\begin{equation}\label{eq:dipole_values1}
\vec{M}_{\uparrow,\Uparrow}=\vec{M}_{\Downarrow,\downarrow}=\vec{0}
\end{equation}
\begin{equation}\label{eq:dipole_values2}
\vec{M}_{\uparrow,\Downarrow(\downarrow,\Uparrow)}=\frac{i}{2}\sqrt{m_0E_p}(1,
(-)i,0)
\end{equation}
where $E_p$ is the bulk material conduction-valence band interaction
energy,\cite{dudi_8kp, pryor} and the spin quantization axis is
chosen along the [001] (or z) direction. For compatibility with the
experimentally defined axes, we choose the major axis of the QD,
believed to be along the [1\=10]
crystallographic axis\cite{gammon} as the x (or H) direction.\\
Substituting Eqs.~(\ref{eq:dipole_momentum}-\ref{eq:dipole_values2})
into Eq.~(\ref{eq:Ehx_Tak}) yields
\begin{widetext}
\begin{equation}\label{eq:Ehx_explicit}
\Delta_1^{n_2,n_1,n_3,n_4}=\frac{3\textsl{e}^2\hbar^2E_p}{2\epsilon
m_0E_g^2}\int\int\phi_h^{n_2*}(\vec{r}_1)\phi_e^{n_1*}(\vec{r}_2)\frac{(y_1-y_2)^2-(x_1-x_2)^2+2i(x_1-x_2)(y_1-y_2)}{((x_1-x_2)^2+(y_1-y_2)^2+(z_1-z_2)^2)^{\frac{5}{2}}}\phi_h^{n_3}(\vec{r}_2)\phi_e^{n_4}(\vec{r}_1)d^3r_1d^3r_2
\end{equation}
\end{widetext}
One can now, in principle, compute these integrals using the single
carrier envelope wavefunctions that were numerically obtained
earlier. This approach demands a lot of computation resources in
order to obtain reliable accuracy. Therefore we choose to
approximate the wavefunctions analytically, by using \mbox{in-plane}
harmonic oscillator functions\cite{glazov2007} instead of the
numerical ones (see Appendix \ref{sect:app:analytic_wavefuncs}).
With these approximations the 6 dimensional integrals are reduced
into 5 analytical ones\cite{eilon_unpublished}. The non-analytical
integral can be easily calculated numerically. Alternatively, for a
nearly round QD, this integral can be expanded into a power series
in the aspect ratio of the model QD, from which only terms up to the
linear order can be kept. This approach provides also important
insight. The result of this derivation for $\Delta_1^{1,1,1,1}$ is
\begin{equation}\label{eq:Ehx_d1111}
\Delta_1^{1,1,1,1}=\frac{3\sqrt{\pi}\textsl{e}^2\hbar^2E_p(\xi-1)}{8\epsilon
m_0E_g^2(l^e_x)^3\xi^2\beta\sqrt{1+\beta^2}}
\end{equation}
where $l_x^e$ is the characteristic length of the electron
(Gaussian) wavefunction in the x direction (see Appendix
\ref{sect:app:analytic_wavefuncs}), $\beta=0.72$ is the ratio
between the characteristic length of the hole wavefunction to that
of the electron,
and $\xi=0.96$ is the length ratio between the short and long sides of the
rectangle (the aspect ratio).\\
For $l_x^e=72$\AA, which gives the same \textit{s-p}$_x$ energy
separation for the electrons as the numerical wavefunctions, we
calculated the $\Delta_1$ terms that we list in Table
\ref{tab:ehx_params}. In the table we also list the values that we
could directly deduce from the measured fine structure splitting of
the X$^0$ and the X$^{-2}$ lines ($\Delta_1^{1,1,1,1}$ and
$\Delta_1^{1,2,1,2}$,
respectively). The agreement, as can be seen in the table, is remarkable.\\
We note that the ratios $\Delta_1^{n_2,n_1,n_3,n_4}$ to
$\Delta_1^{1,1,1,1}$ can be quite large for small deviations from
cylindrical symmetry. In particular, there are significant sign
variations between the various terms. The expressions for these
ratios as functions of $\beta$ and $\xi$ (for $|1-\xi|\ll1$) are
also given in table \ref{tab:ehx_params}.
\begin{table}[tbh]
\caption{\label{tab:ehx_params}Calculated, measured and estimated
electron-hole exchange interaction terms (in $\mu$eV). The
measured and estimated terms were used for calculating the PL
spectra. The calculated ratios are given in terms of the
hole/electron length ratio $\beta$ and the aspect ratio $\xi$, for
$|1-\xi|\ll1$.}
\begin{ruledtabular}
\begin{tabular}{l|c|c|c}
Parameter&Used in fit&Calculated&Ratio to $\Delta_1^{1,1,1,1}$\\
\hline
\hline
$\Delta_0$ & 207 & - & - \\
$\Delta_2$ & 21 & -  & - \\
$\Delta^{1,1,1,1}_1$ & -25 & -15 & - \\
$\Delta^{1,2,1,2}_1$ & 196 & 118 & $\frac{\beta^2}{1+\beta^2}\frac{2\xi-1}{\xi-1}$\\
\hline
$\Delta^{1,3,1,3}_1$ & -222 & -133 & $\frac{\beta^2}{1+\beta^2}\frac{\xi-2}{\xi-1}$\\
$\Delta^{1,4,1,4}_1$ & -6.4 & -3.8 & $\frac{\beta^4}{(1+\beta^2)^2}\frac{41-6\xi}{16}$\\
$\Delta^{1,5,1,5}_1$ & 232 & 139 & $\frac{1}{2}+\frac{\beta^2}{1+\beta^2}\frac{1}{\xi-1}+\frac{\beta^4}{(1+\beta^2)^2}\frac{61\xi-45}{32(\xi-1)}$\\
$\Delta^{2,1,2,1}_1$ & 379 & 227 & $\frac{1}{1+\beta^2}\frac{2\xi-1}{\xi-1}$\\
$\Delta^{2,2,2,2}_1$ & 119 & 71 & $\frac{\beta^2}{(1+\beta^2)^2}\frac{61\xi-45}{16(\xi-1)}$\\
$\Delta^{2,3,2,3}_1$ & -12 & -7.4 & $\frac{\beta^2}{(1+\beta^2)^2}\frac{41-6\xi}{16}$\\
$\Delta^{2,2,2,3}_1$ & 71\textit{i} & 42\textit{i} & $i\frac{\beta^2}{(1+\beta^2)^2}\frac{9-\xi}{16(\xi-1)}$\\
$\Delta^{1,2,1,3}_1$ & 209\textit{i} & 125\textit{i} &
$i\frac{\beta^2}{1+\beta^2}\frac{\xi+1}{2(\xi-1)}$
\end{tabular}
\end{ruledtabular}
\end{table}
\subsection{Optical transitions: Polarization selection rules}
\begin{figure}[tbh]
\includegraphics[width=0.46\textwidth]{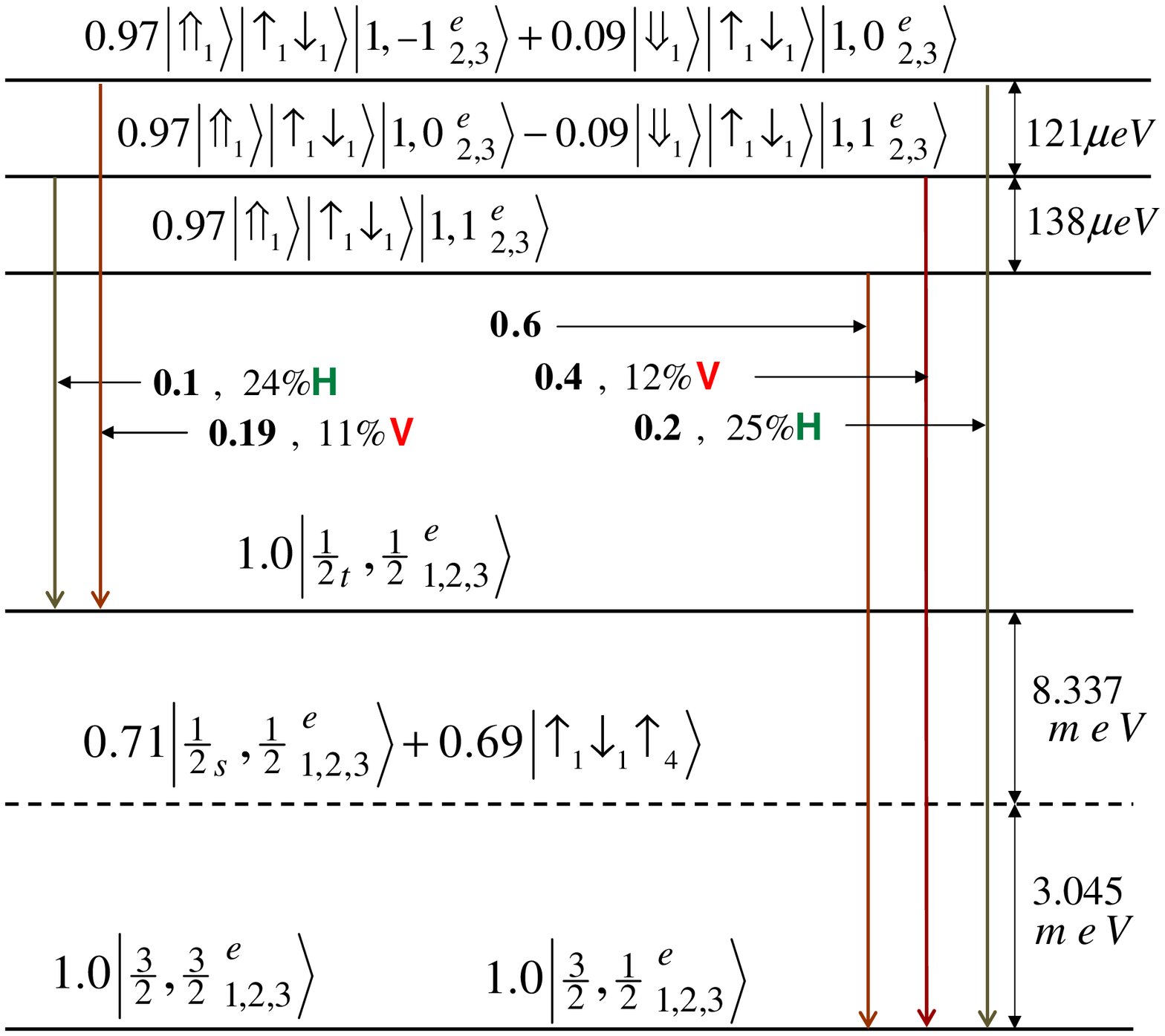}
\caption{\label{fig:x3diagram} Schematic description of the
calculated many carriers energy levels, and their spin
wavefunctions, which optical transition between them result in the
X$^{-3}$ spectral lines. Each transition rate (in units of the total
X$^0$ rate) and its degree of polarization are indicated (if absent
the transition is unpolarized). Only one of the two Kramers states
is shown for each level (for notation see Appendix
\ref{sect:app:symbols}). The number before each component indicates
its amplitude. Components which are irrelevant to the polarization
degree and have amplitudes below 0.1 are not included.}
\end{figure}

\begin{figure}[tbh]
\includegraphics[width=0.5\textwidth]{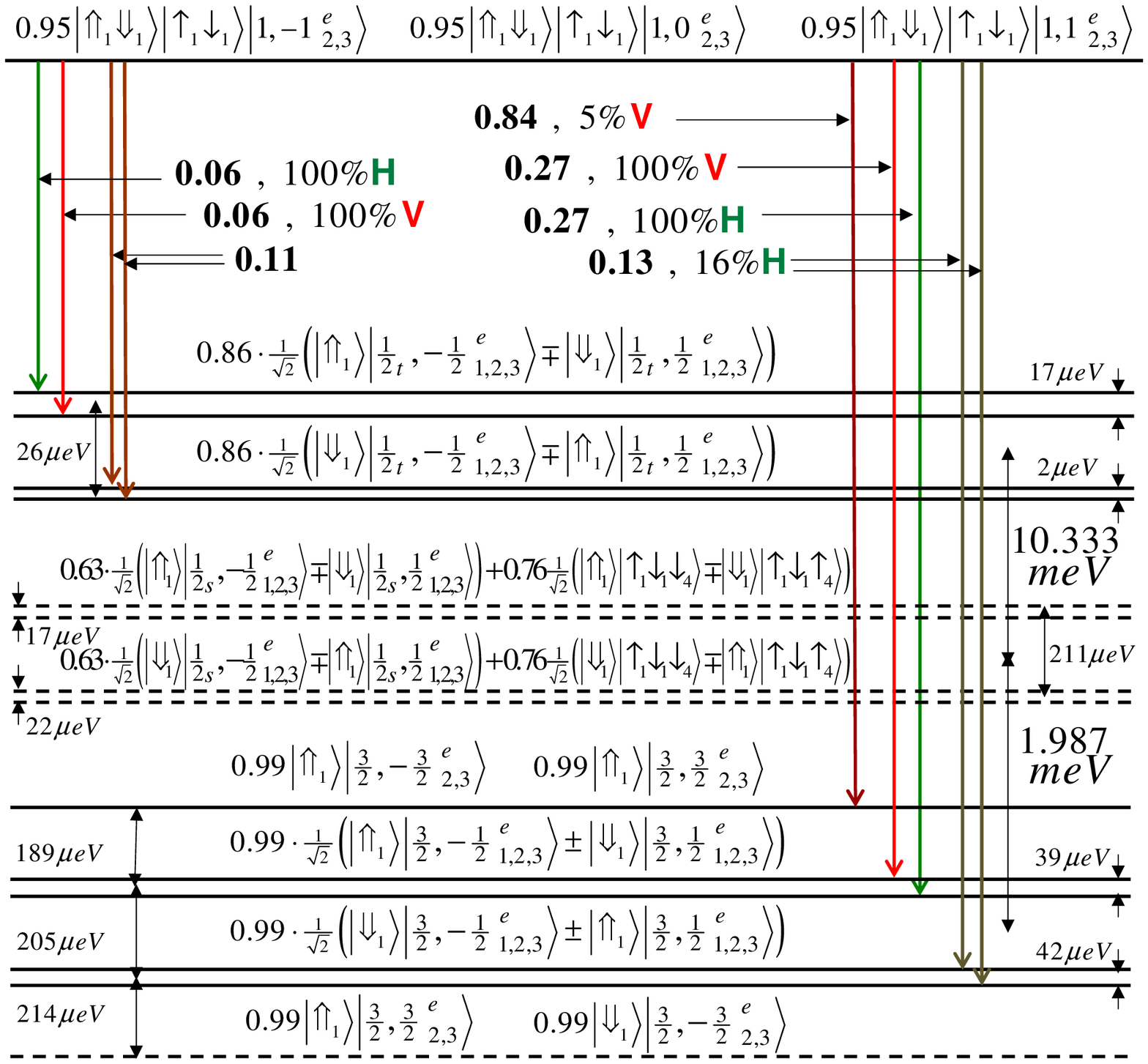}
\caption{\label{fig:xx2diagram} Schematic description of the
calculated many carriers energy levels, and their spin
wavefunctions, which optical transition between them result in the
XX$^{-2}$ spectral lines. Each transition rate (in units of the
total X$^0$ rate) and its degree of polarization are indicated (if
absent the transition is unpolarized). All states are shown (for
notation see Appendix \ref{sect:app:symbols}). The number before
each component indicates its amplitude. Components which have amplitudes
below 0.1 are not included.}
\end{figure}

\begin{figure}[tbh]
\includegraphics[width=0.5\textwidth]{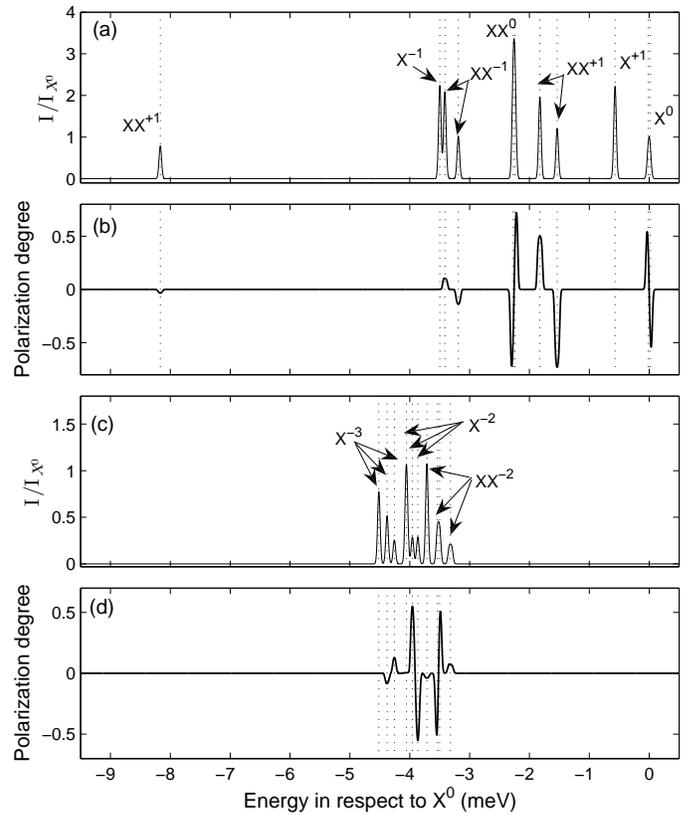}
\caption{\label{fig:calc polar} Calculated PL spectra (a), (c), and
their polarization projections on the H--V axis (b), (d), for
various single QD excitonic transitions. Vertical dash lines at
various spectral lines are drawn to guide the eye.}
\end{figure}

The optical transition operator in the dipole approximation is
expressed as:\cite{barenco}
\begin{equation}\label{eq:transition_op}
\hat{\vec{P}}=\sum_{i,j}{\vec{p}_{ij}\hat{a}_i\hat{b}_j}
\end{equation}
Under the one-band SVEFA, the transition momentum vector $\vec{p}_{ij}$ is
given by:
\begin{equation}\label{eq:dipole_vec_env}
\vec{p}_{ij}=\vec{M}_{S_j,S_i}\int{\phi^{e*}_i(\vec{r})\phi^{h}_j(\vec{r})d^3r}
\end{equation}
The momentum matrix elements $\vec{M}_{S_j,S_i}$ are given
explicitly by
Eqs.~(\ref{eq:dipole_values1}-\ref{eq:dipole_values2}).\\
The rate of an optical transition\cite{henry1970} centered at an
energy $\varepsilon$, for a certain polarization $\vec{e}$ is given
by:
\begin{equation}\label{eq:polarized power}
\Gamma_{\vec{e}}(\varepsilon)=\frac{4\alpha n\varepsilon}{3\hbar
m_0^2c^2}\sum_{i,f}{|\langle
f|\vec{e}\cdot\hat{\vec{P}}|i\rangle|^2\delta_{\varepsilon,\varepsilon_i-\varepsilon_f}F_i}
\end{equation}
where $\alpha=\frac{\textsl{e}^2}{\hbar c}\approx\frac{1}{137}$ is
the fine structure constant, and $n$ is the refraction index of the
QD material. The indices $i$ and $f$ run over all initial states
$|i\rangle$ and final states $|f\rangle$. $\varepsilon_i$
($\varepsilon_f$) is the energy of the initial state $|i\rangle$
(final state $|f\rangle$). $F_i$ is the population probability of
the initial state $|i\rangle$. For the bright neutral exciton
transitions we calculate
\begin{equation}\label{eq:x0_matrix_elements}
|\langle 0|\hat{x}\cdot\hat{\vec{P}}|X^0_H\rangle|^2=|\langle
0|\hat{y}\cdot\hat{\vec{P}}|X^0_V\rangle|^2=\frac{m_0E_p}{2}\cdot1.44
\end{equation}
The other two transitions from the bright states and the transitions
from the `dark' excitonic states, vanish. Assuming equal population
probabilities for the bright and dark X$^0$ states, we get a total
X$^0$ rate of (0.78ns)$^{-1}$, in agreement with the measured
lifetime\cite{nika_ent}. The calculated rates of all other optical
transitions
are given in units of this total X$^0$ rate.\\
For example, in Fig.~\ref{fig:x3diagram} we present a diagram of the
calculated many carriers energy levels and the optical transitions
between them, which lead to the spectrum resulted from excitonic
transitions in a triply negatively charged QD (X$^{-3}$).\\
As can be deduced from Fig.~\ref{fig:x3diagram}, the X$^{-3}$ line
results from three initial levels (each doubly Kramres degenerate).
These levels contain mainly the following single carrier states: one
s-shell hole, two s-shell electrons one $p_x$ and one $p_y$
electron, where the $p$-shell electrons are in their triplet
configurations. These open shells configurations are the lowest in
energy, since the energy difference between the $p_x$ and $p_y$
single electron states is smaller than their exchange interaction.
The degeneracy between the triplet configurations is removed by the
EHEI with the hole. Our experimental data can only be explained with
these open p-shells
occupation in mind (see below).\\
The final states are mainly composed of three single electron
states: one in the s-shell, one in the $p_x$ shell and one in the
$p_y$ shell. The expected eight fold degeneracy is partially
removed by the electron - electron exchange interaction, which
leave a four fold  degenerate ground state (we found no
experimental evidence for an anisotropic e-e exchange interaction
\cite{ware} which would have further reduced this degeneracy). The
calculated optically allowed transitions between the initial and
the final states and their polarization selection rules are given
in Fig.~\ref{fig:x3diagram}. The highest energy transition is
finely structured from three lines with intensity ratios of
approximately 3:2:1. These intensities were previously deduced
using a simple model, by Urbaszeck et al \cite{urbazek}.\\
If the ground p-shell was occupied by two electrons, the X$^{-3}$
transitions would have generate a single almost unpolarized spectral
line, very similar to that due to the X$^{-1}$ transitions. This is
in clear contradiction with the measurements presented in
Fig.~\ref{fig:PL_polar}, Fig.~\ref{fig:finestruct_comp} and previous
measurements on similar QDs\cite{urbazek,warburton_zunger}.\\
Another example is provided in Fig.~\ref{fig:xx2diagram} where we
show the levels' diagram and optical transitions, which result in
the biexciton recombination in a doubly negatively charged QD
(XX$^{-2}$).\\
Here the initial states are mainly composed of the same single
carrier states as the initial states of the X$^{-3}$, except for the
addition of one more s-shell hole. Unlike the X$^{-3}$, the paired
s-shell holes do not remove the degeneracy of the triplet
configuration of the p-shell electrons. Similarly, the final states
are mainly composed of the same single electron states as the final
states of the X$^{-3}$, except for an additional one s-shell hole.
The EHEI between the unpaired hole and the electrons, completely
removes the degeneracy between the electron states. As a result,
there are eight low energy states (the energy differences between
the lowest and between the highest pairs of states are too small to
be noted) to six of which optical transitions are allowed. Similar
to the case of the X$^{-3}$, the optical spectrum is finely
structured from three pairs of lines with total intensity ratios of
approximately 3:2:1 as previously deduced by the simple
considerations of Urbaszeck et al \cite{urbazek}. A major difference
between the two examples is in their polarization selection rules.
In the first case (X$^{-3}$) the total spin is half integer and
Kramers degeneracy prevails. Therefore only partial linear
polarization is expected. In the second case (XX$^{-2}$) the total
spin is an integer. In this case, full linear polarization is
expected, just like in the case of the neutral single exciton.
Indeed, the intermediate pair of spectral lines is fully polarized
along
the QD's primary axes.\\
Similar diagrams for less complicated transitions were discussed in
previous works\cite{bayer_rev,akimov,kalevich}. In these works
simpler models were used. These models are sufficient only when the
EHEIs are much smaller than any other interaction.
\\
In Figs. \ref{fig:calc polar}(a) and \ref{fig:calc polar}(c) we show
the calculated spectrum for various charge states. The corresponding
H--V polarization projections are shown in Figs. \ref{fig:calc
polar}(b) and \ref{fig:calc polar}(d), respectively.\\
Within our simple, one band model, the calculated D-\=D (and of
course the R--L) projections vanish, and therefore they are not
shown. For the calculations, equal probabilities for excitons and
biexcitons in all charge states were assumed~\cite{dekelssc}. In the
calculations, only initial configurations within 1~meV (compatible
with the experiments' temperature) above the ground state were
considered.\\
The calculated lines are convoluted with 50$\mu$eV broad Gaussians,
to account for the spectral diffusion\cite{nika_ent}. In the
calculation of the polarization projections, a constant background
of 3.5\% of the maximal intensity is added to both cross-linearly
polarized spectra. This is done in order to mimic the effect of
background noise on the measured spectra (see sect. \ref{sec:exp}).
\section{\label{sec:comp}
Comparison between experimental measurements and model calculations}
\begin{figure}[tbh]
\includegraphics[width=0.45\textwidth]{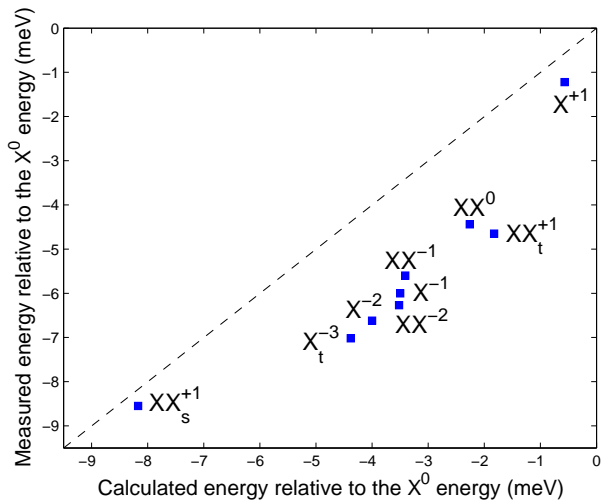}
\caption{\label{fig:comp} A comparison between measured and
calculated spectral positions of a few lines. The dashed line is the
equality line. The size of the markers represent the experimental
error.}\end{figure}

\begin{figure}[tbh]
\includegraphics[width=0.5\textwidth]{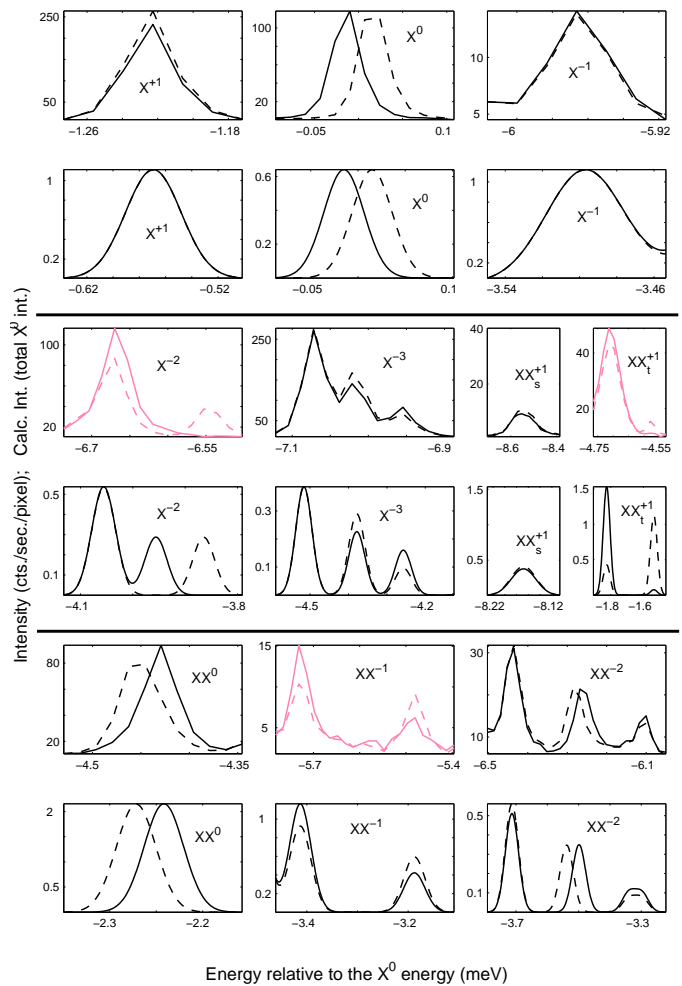}
\caption{\label{fig:finestruct_comp} (color online) Measured (top
panel in each pair) and calculated (bottom panels) high resolution
polarization sensitive PL spectra of various spectral lines. The
solid (dashed) black line represents H (V) polarized spectrum while
the solid (dashed) pink line represents the V+\=D (H+D) polarized
spectrum.}\end{figure}

\begin{figure}[tbh]
\includegraphics[width=0.5\textwidth]{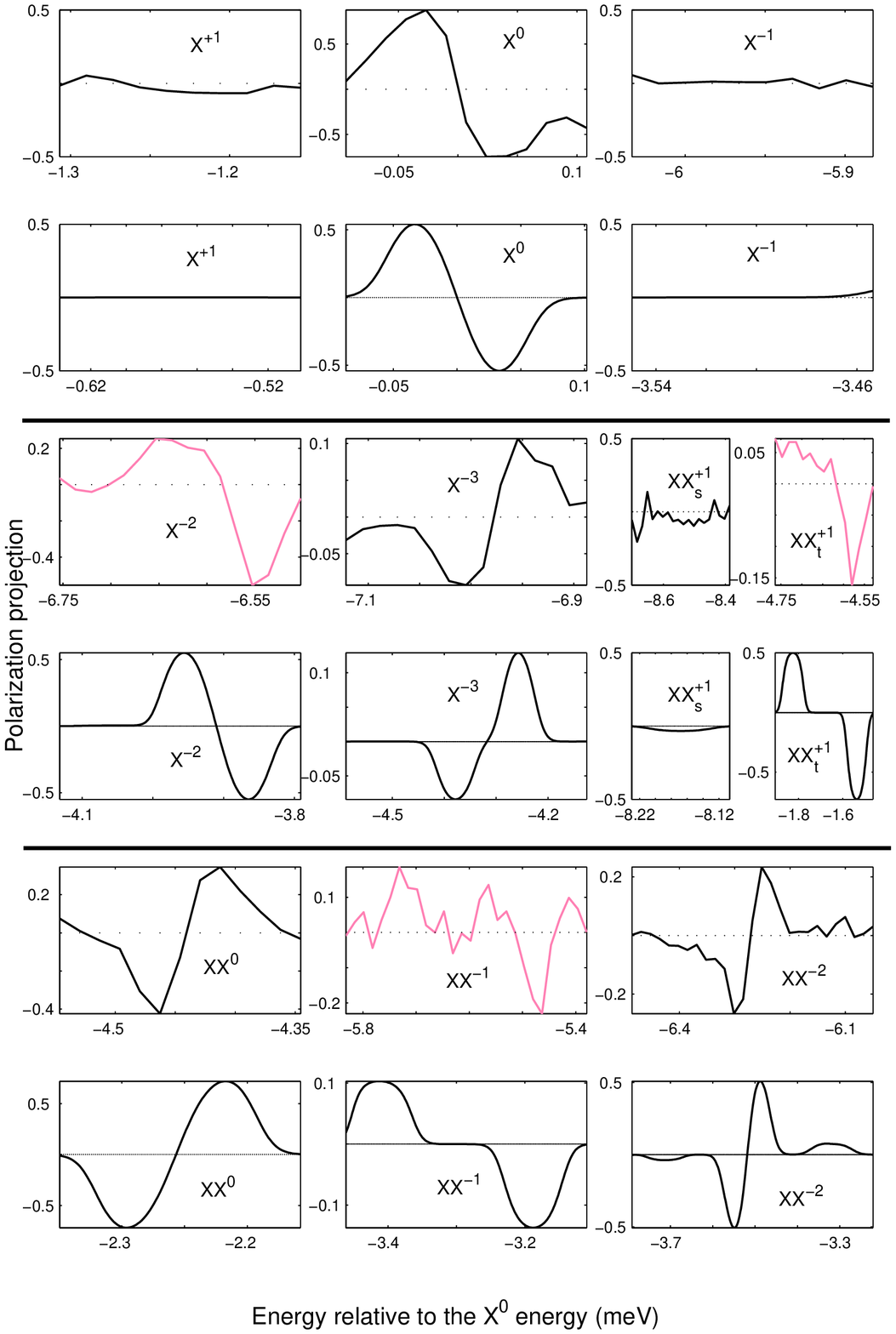}
\caption{\label{fig:polproj_comp} (color online) Measured (top
panels) and calculated (bottom panels) linear polarization spectra
of various spectral lines. Black (pink) solid lines  represent
projections along the H--V ((V+\=D)--(H+D)) axis of the poincar\'e
sphere.}
\end{figure}

In Fig.~\ref{fig:comp} we compare between the measured and
calculated spectral positions of various lines.
\\
We note that the spectrum produced by our
simple model correlates with the experimentally measured one in the
energy order of the various spectral lines. The calculated energy
differences between the various lines, however, are in most cases
smaller than those measured. Specifically, the calculated energy
differences between the exciton and the biexciton and between the
positive and negative trions (X$^{+1}$ and X$^{-1}$, respectively)
are smaller than the measured ones. This is probably a consequence
of the relative simplicity of our single band model\cite{regelman2}
and the lack of information about the exact shape strain and
composition of the QDs. With our model's limitations we found it
hard to simultaneously fit the biexciton binding energy and the
difference between the positive
and negative trion transitions.\\
In Fig.~\ref{fig:finestruct_comp} we compare the measured and
calculated polarized fine structure of various spectral lines, while
In Fig.~\ref{fig:polproj_comp} we compare the measured and
calculated linear polarization spectra for these spectral lines.
\\
We note in
Figs.~\ref{fig:finestruct_comp}~and~\ref{fig:polproj_comp} that the
measured fine structures are reproduced quite nicely by our model
calculations. In particular, the calculated number of fine structure
components, their relative intensities and their polarizations
correlate with the measured
values.\\
On the other hand, while the calculated polarization spectra are
always polarized along the \mbox{H--V} axis of the poincar\'e
sphere, the measured ones are sometimes rotated. Few specific lines
(see below) are polarized along the \mbox{(V+\=D)--(H+D)} axis.\\
The calculated energy differences between the fine structure
components of a given spectral line are sometimes larger than the
measured values. Particularly, the calculated fine structure
splittings between the components of the X$^{-3}$ line, and the
calculated splitting between the unpolarized and polarized doublets
of the X$^{-2}$ line are larger than the measured ones. We believe
that this maybe a consequence of the $\Delta_0$ dependencies on the
envelope wave functions, which are neglected in our model.\\
In the absence of polarization memory (which requires quasi-resonant
polarized excitation \cite{ware}), and in the absence of magnetic
field, the theory predicts that the spectral lines can only present
linear polarizations. This is what we observed experimentally as
well. In the theoretical model, the linear polarization can only be
oriented along the main axes of the QD, which are usually along the
crystalline directions [1\=10] and [110] \cite{gammon}, which we
denote by H and V respectively. In the experiment, however, we found
that some spectral lines are polarized along other directions. The
polarization appears in \textit{three} sets of orthogonal axes: The
measured polarization of the neutral exciton line is indeed along
the H--V axis of the Poincar\'e sphere. A few other lines, notably
the neutral biexciton, the doubly charged biexciton and the triply
charged exciton, are also polarized along this axis. Their degree of
polarization is somewhat smaller than that of the X$^0$ lines due to
the unpolarized spectral background that they ride on. Few other
spectral lines  are polarized along an axis which is rotated
clockwise by 135 degrees relative to the H--V axis of the
Poincar\'e sphere.\\
This polarization axis, which roughly coincides with the
$\frac{1}{\sqrt{2}}$(V+\=D) ([120]) and the
$\frac{1}{\sqrt{2}}$(H+D) ([2\=10]) crystalline directions,
appears only in lines associated with configurations which contain
one unpaired $p_x$ carrier (either electron or hole): X$^{-2}$,
XX$^{-1}$ and XX$^{+1}$. Careful inspection of the unevenly
polarized spectra of the X$^{-2}$ line presented by Ediger et
al\cite{warburton_zunger} leads to the same conclusion.
Unfortunately, they did not fully measured the polarization state of the line.\\
For other configurations, which contain only $s$ carriers or either
closed shells or two unpaired $p$ carriers ($p_x$ and $p_y$), the
polarization is along the H--V axis. In addition, we sometimes
observe lines which are polarized along the D--\=D axis as well.
Such a spectral line is seen in Fig.~\ref{fig:PL_polar}(d), at
energy of -2.6 meV relative to the X$^0$ line. This relatively weak
doublet, may result from pair recombination in doubly negatively
charged QD as deduced
from its voltage dependence.\\
These novel observations are not reproduced by the simple model that
we present, probably since the model ignores the underlying crystal.
For example, in the single band model the polarization due to
recombination of a $p_x$ shell e-h pair is the
same as that of a $p_y$ pair, which is clearly not the case.\\
We rule out another possible explanation\cite{ivchenko2005} expected
in a nearly \textit{p}-shell degenerated QD. In such a QD, where the
$p_x$-$p_y$ splitting is comparable to or smaller than the EHEI,
configurations containing an unpaired $p_x$ carrier and those
containing an unpaired $p_y$ carrier are mixed\cite{ivchenko2005}.
This mixing can indeed lead to recombination in linear polarizations
along axes different than the primary axes of the QD. This is not
the case here, since for nearly degenerate QDs, the fine structures
of the X$^{-2}$, XX$^{-1}$ and XX$^{+1}$ transitions should contain
twice the number of fine structure components than that actually
observed.\\
One straight forward way to include the lattice will be to solve the
single electron problem using a multi-band approach\cite{dudi_8kp,
hawrylak}.\\
Such a model is absolutely necessary for calculating the
polarization selection rules in highly positively charged QDs.
There, the complicated nature of the p-shell holes is not likely to
be captured by a one band model. The polarization degree of the
highly positively charged QD emission lines (identified by their
voltage dependence) that we measured was marginal, while our single
band model yields polarizations similar to those of the negatively
charged QD lines. We believe that this discrepancy results from the
inadequacy of the single band model. In addition, we note that the
highly positively charged QD lines were all measured under large
electrostatic fields, which for now, were not considered in our
model.
\section{\label{sec:sum}Summary}
We presented detailed polarization sensitive spectroscopy of single
QDs in various charge states. We developed a many-carrier model
based on single band, envelope function approximation, which
includes the isotropic and anisotropic electron-hole exchange
interactions for the analysis of the measured data. We calculated
the PL spectrum with its fine structure and polarizations for the
exciton and biexciton optical transitions in neutral, singly
positively and singly, doubly and triply negatively charged QDs. The
calculations are favorably compared with the measured polarization
sensitive PL spectra.\\
However, while our model can only reproduce polarizations along the
main axes of the QD, the experimental data display polarizations
oriented along other directions as well. This indicates, probably,
that the one band based model is too simple to describe this novel
observation.\\
\begin{acknowledgments}
The research was supported by the US-Israel Binational Science
Foundation (BSF), the Israeli Science Foundation (ISF) and the
Russell Berrie Nanotechnology Institute (RBNI) at the Technion. We
acknowledge fruitful correspondence with professor E.L. Ivchenko.
\end{acknowledgments}
\appendix
\section{\label{sect:app:analytic_wavefuncs}The analytical wavefunctions used in the calculation of the $\Delta_1$ integrals}
As model functions we used the following functions:
\begin{center}
\mbox{$\phi^p_1=|0,0\rangle_p$ ; $\phi^p_2=|1,0\rangle_p$ ; $\phi^p_3=|0,1\rangle_p$}\\
\end{center}
\begin{center}
\mbox{$\phi^p_4=|1,1\rangle_p$ ; $\phi^p_5=|2,0\rangle_p$ ; $\phi^p_6=|0,2\rangle_p$}\\
\end{center}

The kets of the form $|n_x,n_y\rangle_p$ stand for the 2D elliptic
harmonic oscillator functions:
\begin{equation*}\label{eq:HO_wf}
\langle
x,y|n_x,n_y\rangle_p=\frac{H_{n_x}(\frac{x}{l_p^x})H_{n_y}(\frac{y}{l_p^y})}{\sqrt{2^{(n_x+n_y)}n_x!n_y!\pi
l_p^xl_p^y}}e^{-\frac{1}{2}((\frac{x}{l_p^x})^2+(\frac{y}{l_p^y})^2)}
\end{equation*}
where $p$ is the charge carrier index (either `$e$' or '$h$'),
$l_p^{x(y)}$ is a characteristic length along the $x$($y$)
direction, $n_{x(y)}$ is the quantum number associated with the
$x$($y$) direction, and $H_{n_{x(y)}}$ is the Hermite polynomial of
order $n_{x(y)}$. The aspect ratio $\xi$ and the hole/electron
length ratio $\beta$ are defined as:
\mbox{$\xi=\frac{l_e^y}{l_e^x}=\frac{l_h^y}{l_h^x}$} ;
\mbox{$\beta=\frac{l_h^x}{l_e^x}=\frac{l_h^y}{l_e^y}$}.
\section{\label{sect:app:symbols}The notation used for the state vectors}
In table \ref{tab:state_symbols} we list the symbols and
abbreviations used in writing the spin state vectors.
\begin{table}[tbh]\caption{\label{tab:state_symbols}
The abbreviations used to write the state vectors. $\uparrow_j$
($\Downarrow_j$) stands for an electron (hole) with
envelope-function number $j$ and pseudo-spin up (down). $j=1$ is the
$s$ state, $j=2$ is the $p_x$ state, $j=3$ is the $p_y$ state, $j=4$
is the $d_{xy}$ state, $j=5$ is the $d_{xx}$ state, and $j=6$ is the
$d_{yy}$ state. Where two or more of \textit{i,j} and \textit{k}
appear together, it is assumed that \mbox{$i\neq j\neq k$}.}
\begin{ruledtabular}
\begin{tabular}{l|r}
Abbreviation & Full form \\
\hline
\hline
$|0,0\ _{ij}^e\rangle$ &  $\frac{1}{\sqrt{2}}(|\uparrow_i\downarrow_j\rangle-|\downarrow_i\uparrow_j\rangle)$ \\
$|1,0\ _{ij}^e\rangle$ &  $\frac{1}{\sqrt{2}}(|\uparrow_i\downarrow_j\rangle+|\downarrow_i\uparrow_j\rangle)$ \\
$|1,1\ _{ij}^e\rangle$ &  $|\uparrow_i\uparrow_j\rangle$\\
$|1,$-$1\ _{ij}^e\rangle$ &  $|\downarrow_i\downarrow_j\rangle$\\
$|\frac{3}{2},\frac{1}{2}\ _{ijk}^e\rangle$ &  $\frac{1}{\sqrt{3}}(|\uparrow_i\uparrow_j\downarrow_k\rangle+  |\uparrow_i\downarrow_j\uparrow_k\rangle+|\downarrow_i\uparrow_j\uparrow_k\rangle)$\\
$|\frac{3}{2},$-$\frac{1}{2}\ _{ijk}^e\rangle$ &  $\frac{1}{\sqrt{3}}(|\downarrow_i\downarrow_j\uparrow_k\rangle+  |\downarrow_i\uparrow_j\downarrow_k\rangle+|\uparrow_i\downarrow_j\downarrow_k\rangle)$\\
$|\frac{3}{2},\frac{3}{2}\ _{ijk}^e\rangle$ &  $|\uparrow_i\uparrow_j\uparrow_k\rangle$\\$|\frac{3}{2},$-$\frac{3}{2}\ _{ijk}^e\rangle$ &  $|\downarrow_i\downarrow_j\downarrow_k\rangle$\\
$|\frac{1}{2}_t,\frac{1}{2}\ _{ijk}^e\rangle$ & $\sqrt{\frac{2}{3}}|\uparrow_i\uparrow_j\downarrow_k\rangle-\frac{1}{\sqrt{6}}|\uparrow_i\downarrow_j\uparrow_k\rangle+\frac{1}{\sqrt{6}}|\downarrow_i\uparrow_j\uparrow_k\rangle$\\
$|\frac{1}{2}_t,$-$\frac{1}{2}\ _{ijk}^e\rangle$ &  $\sqrt{\frac{2}{3}}|\uparrow_i\uparrow_j\downarrow_k\rangle+\frac{1}{\sqrt{6}}|\downarrow_i\uparrow_j\downarrow_k\rangle-\frac{1}{\sqrt{6}}|\uparrow_i\downarrow_j\downarrow_k\rangle$\\
$|\frac{1}{2}_s,\frac{1}{2}\ _{ijk}^e\rangle$ &  $\frac{1}{\sqrt{2}}(|\uparrow_i\uparrow_j\downarrow_k\rangle-|\uparrow_i\downarrow_j\uparrow_k\rangle)$\\
$|\frac{1}{2}_s,$-$\frac{1}{2}\ _{ijk}^e\rangle$ &  $\frac{1}{\sqrt{2}}(|\downarrow_i\downarrow_j\uparrow_k\rangle-|\downarrow_i\uparrow_j\downarrow_k\rangle)$\\
\end{tabular}
\end{ruledtabular}
\end{table}
\bibliography{epdg07}
\end{document}